# The Many Worlds of Hugh Everett III

Adrian Kent

Centre for Quantum Information and Foundations, DAMTP, University of Cambridge, Wilberforce Road, Cambridge CB3 0WA, U.K.

*and*

Perimeter Institute for Theoretical Physics, 31 Caroline Street North, Waterloo, Ontario, Canada  N2L 2Y5.

**Abstract**  A review of Peter Byrne's biography [1] of Hugh Everett III, to appear in the American Journal of Physics.

Hugh Everett III occupies a peculiar place in the history of physics, famous for his sole contribution, the idea that quantum theory can best be understood as describing many equally real alternative worlds corresponding to the different possible classical outcomes arising from quantum events.   It is an idea which Everett himself was never able properly to develop, which remains ill-understood, and may indeed not even make rigorous sense, but which nonetheless has greatly influenced modern developments in theoretical physics.

In a romantic world, Everett might have struggled through hardship to develop and champion his version of quantum theory, and emerged ultimately vindicated -- perhaps even today giving colloquia and inspiring younger generations to ask fundamental questions and pursue their ideas despite setbacks.  In prosaic fact, although encouraged and supported by his PhD supervisor, John Archibald Wheeler, and offered opportunities to pursue a good academic career, he unromantically chose to spend the rest of his life making a mostly comfortable living as a software developer and military consultant, focussing particularly on developing optimal strategies for the US to deter and/or fight a large-scale nuclear war.    He published nothing on quantum theory beyond his PhD work.  While he was gratified when his ideas eventually began receiving wider attention, he seems to have cared surprisingly little about either the fate or the ultimate validity of his view of quantum theory.  He died sadly early, in 1982, aged 51, of a heart attack to which his chain smoking, alcoholism, fondness for rich food and depression most likely contributed, leaving his wife, Nancy, with instructions to throw his ashes in the trash (which she eventually did) and two children, Liz and Mark, to whom he was more enigma than parent.

David Deutsch, one of the founders of the theory of quantum computing -- and one younger physicist who definitely was directly inspired by Everett, through lectures and conversations at the University of Austin in 1977 -- is quoted on the back cover as describing Everett as one of the finest minds of the twentieth century.   One of the virtues of Peter Byrne's thoughtful biography is that it refrains from this sort of crude intellectual mythology.   On the contrary, Byrne is a refreshingly unhagiographic populariser: he sees the intellectual history of many-worlds quantum theory as often determined more by the complex agendas and personal loyalties of the participants than by scientific or logical argument, and Everett, Wheeler, Petersen, Rosenfeld and even (whisper it) Bohr emerge as sometimes confused actors, as well as less than saintly ones.

Byrne's account reinforced my impression from Everett's two published papers on quantum theory that, judged by the very highest scientific standards, he was actually a somewhat impatient and un-self-critical thinker.   He groped around the key questions of how (if) distinct worlds can be said to emerge from a universal wave function, and what (if anything) probability can mean in such a picture, without ever carefully addressing them or acknowledging the large gaps in his arguments that others pointed out and that many theorists sympathetic to his ideas have subsequently tried to fill.   One of the remarkable discoveries Byrne reports is  Everett's personal copy of DeWitt and Graham's 1973 volume, "The Many Worlds Interpretation of Quantum Mechanics", found by a correspondent in a second-hand bookshop.    It contains some notably intemperate annotations -- "bullshit", "Goddamit (sic) you don't see it", and so forth -- scrawled on the work of some of his leading *supporters.*

Perhaps one should not read too much into the possibly alcoholic marginalia of the later Everett, but they do fit into a lifelong pattern of resistance to thoughtful criticism.  Like a lazy student who knows the solution he is supposed to derive but struggles to justify it, Everett  had a convenient tendency to conflate difficult questions about quantum theory with easier ones.   Instead of explaining how we can derive the appearance of a single world following standard quantum probabilistic laws from many-worlds quantum theory, he offered a proof that a probability-like function defined on branching worlds must take the familiar form of the Born rule if it satisfies some mathematically appealing conditions -- not a completely trivial question, to be sure, but much easier than, and only tangentially relevant to, the one he claimed to be solving.   Nor did he tackle the key question of how to justify  basing an interpretation on a very particular decomposition of the universal wave function, picking out a basis in which it takes the form of a superposition of many worlds like ours which are well approximated by classical theories and contain creatures like us that perceive and exploit this classical predictability.    He simply took this preferred basis decomposition as given and then analysed how memory record states would be appropriately correlated with one another and with environment states -- indeed a sensible point to check if one can resolve the preferred basis problem, but not a substitute for a solution.

Everett's ideas here seem, incidentally, to have been quite different from those of his modern supporters.  Byrne digs out an intriguing 1962 conference transcript recording Everett and Podolsky agreeing that the number of distinct worlds in the universal wave function should be not just infinite, but uncountably infinite.   Everett here seems to take literally the picture outlined in a toy model in his thesis, in which literally every possible value of the position coordinate of any given measured particle corresponds to a distinct world, whereas modern Everettians mostly envisage a large finite or countably infinite number of distinct (albeit, they say, fuzzily defined) worlds defined (they argue) by the physics of decoherence.

Byrne gives us some fascinating and illuminating descriptions of Everett's thesis work, Wheeler's influence on Everett's presentation, and their subsequent unsuccessful attempts to interest Petersen, Bohr and the Copenhagen school in Everett's ideas, of Everett's occasional later forays into academic debate, and his eventual championing by DeWitt, Deutsch and others.    Alongside this, the book gives a vivid account of what became Everett's main career, beginning from work on game theory at Princeton and becoming a significant player as a military-industrial-governmental Cold War strategist.   Byrne uses the life stories of Nancy and Hugh Everett to reconstruct the social and political climate of that era.   There is much fascinating, if chilling, material in these chapters on the uses and abuses of game theory in war planning, the inhuman but seemingly inescapable logic that led to the development of mutual assured destruction as an official strategic goal, and the bloodthirsty psychopathy of Herman Kahn and some other influential American strategists (surely, it should be said, more than matched by their Russian counterparts, although this comparison goes beyond Byrne's scope).

One might perhaps imagine a biographer with Byrne's reportorial curriculum vitae, which includes contributions to Mother Jones, SF Weekly and the North Bay Bohemian, could end up depicting Everett as a monster.   His flaws are indeed brought out clearly -- perhaps most of all, his sad and disturbing detachment from both humane and scientific values.   Everett wanted a PhD thesis from Princeton, and the idea of a many-worlds interpretation became his route to one.   He wanted a career that would give him a financially comfortable lifestyle: Cold War strategist fitted the bill.   It is hard to find evidence in this biography that he ever really cared much about advancing our understanding of nature, the outcome of the Cold War, the fate of the tens of millions whose lives were variables in his calculations, the happiness of his family, or, in fact, anything at all.

Yet, as Byrne suggests in his introduction, the book also often inclines towards understanding and forgiveness.   And credit is due: Everett's thesis work did offer a completely fresh perspective on quantum theory, which the greatest contemporary physicists were mostly too rigidly dogmatic or unimaginative to appreciate, and so failed to respond adequately to.   As Byrne's fascinating extracts from correspondence and dialogues between Everett, Wheeler and the Copenhagen school vividly illustrate, making sense of quantum theory is, unfortunately, something that human beings are just not very good at.   We seem to be too easily driven by the discomfort of uncertainty and puzzlement to seize on partial truths and incomplete ideas and stiffen them into articles of faith.   It seems unlikely to me that Everett's vision of many-worlds quantum theory will ever be made into a coherent scientific theory capable of explaining experimental data, but it *was* unquestionably liberating in helping to take theoretical physics beyond Copenhagen doctrine, and I think it will ultimately be seen as thus having significantly contributed to advancing our understanding of quantum theory and of nature.   Most of us would be happy enough with such a scientific epitaph.

Everett's human failings should also be seen in perspective.   They were clearly many, but not egregiously uncommon.   He was not the first or last scientist to treat science as a game, or to abandon pure science for a materially comfortable career of questionable social value.   Nor does he seem to have been a particularly atypical Cold War careerist.   And, terrifying though some of his and his colleagues' attitudes and calculations were, one has to be adult about the nature of their work: planning and preparation for possible nuclear war were almost inevitable given post-1945 geopolitics, and many better scientists with stronger consciences -- and in some cases more idealistic motivations -- than Everett worked to develop weapons and strategies.   If the experience of World War II and the Cold War taught us anything about how to avoid destroying ourselves with nuclear or biological weapons in future, it is surely to focus on the underlying dynamics of conflicts rather than hoping we will be saved by miraculously aligned mass moral decisions to resist among the scientists on both sides.

Sadly, too, Everett was not so very unusual in having found it hard to form close human relationships -- a difficulty to which his family background and the nature of his work both probably contributed.   Everett's later personal and professional lives show how much harm a game-theoretic model of life, devoid of humane values, can cause, and maybe the main interest of this part of his story is in reinforcing awareness of this vulnerability.   His son, Mark Oliver Everett, records in a foreword coming to some understanding for, forgiveness of, and identification with his father through helping Byrne's work on this biography, and it is easy to respect these feelings.  Perhaps it is more appropriate to feel sorry for Everett, and to try to learn lessons from his life, than to judge.

Byrne has not written -- or I think tried to write -- a definitive intellectual history of many-worlds quantum theory, nor a completely authoritative biography.   He superimposes his own judgements, personal and political, a few of which jar: for example "Pure science was not, in Einstein's view, an end in itself." (p. 10), or the perhaps more plausible, but to my mind over-confidently reductive, declaration that "[Everett's] inability to resolve his ambivalent feelings toward [his mother] festered, causing him to distrust humanity for reasons he could not fully explain." (p. 24)   The book is also blemished by Oxford University Press' unforgivably careless copy editing.   While occasionally this amuses -- the surreal image of discomforted heteronormative eels conjured up by "conventional sexual morays" on p. 44 was a highlight -- cumulatively the misspellings and other solecisms grate.   If publishers want to preserve a niche for that endangered species, the printed book, they need to take much greater care in their work.

Byrne's narrative nonetheless compels serious attention, contains much important new material, is greatly enlivened and enhanced by his eagle eye for the telling quotation, and is always interesting and often convincing.  It should intrigue any student of twentieth century physics, and is also a valuable resource for anyone concerned with the broader education of scientists and the impact narrowly scientific ways of thinking can have on scientists themselves and on the wider world.

Adrian Kent is Reader in Quantum Physics at the Department of Applied Mathematics and Theoretical Physics, University of Cambridge and an Associate Faculty member of Perimeter Institute for Theoretical Physics.   He co-edited "Many Worlds?  Everett, Quantum Theory and Reality" (Oxford University Press, 2010).